\documentclass[a4paper,11pt]{article}
\pdfoutput=1
\usepackage{jcappub} % for details of the package, see JCAP-author-manual
\usepackage[T1]{fontenc} % if needed
\usepackage{graphics}  
\usepackage{graphicx}                     \usepackage{natbib}

\usepackage{color}                                                              
\usepackage{ifthen}                                  %\usepackage{hyperref}
\usepackage{mathrsfs}
\usepackage{amsmath, amssymb}
\usepackage{multirow}                                                          
\usepackage{tabularx}
\usepackage{subcaption} 
\usepackage{titlesec}
\usepackage{array}% 
\usepackage{float}

\def\nn{\nonumber}

\def\da{{\rm d}a}

\def\dl{{\rm d}\ell}
\def\d{{\rm d}}
\def\S2{\mathcal{S}^2}
\begin{document}

\title{The nature of non-Gaussianity and statistical isotropy of the 408 MHz Haslam synchrotron map}
\author[a,b,1]{Fazlu Rahman}{\note{corresponding author}}
\author[a]{Pravabati Chingangbam}
\author[c]{Tuhin Ghosh}

\affiliation[a]{Indian Institute of Astrophysics, Koramangala II Block,       
  Bangalore  560 034, India}
\affiliation[b]{Department of Physics, Pondicherry University, R.V. Nagar, Kalapet, 605 014, Puducherry, India}
\affiliation[c]{National Institute of Science Education and Research Bhubaneswar, P.O. Jatni, Khurda 752 050, Odisha, India}  

\emailAdd{fazlu.rahman@iiap.res.in}
\emailAdd{prava@iiap.res.in}
\emailAdd{tghosh@niser.ac.in}

%%%%%%%%%%%%%%%%%%%%%%%%%%%%%%%%%%%%%%%%%%%%%%%%%%%%%%%%%%%%%%%%%%%%%%%%%%%
\abstract{Accurate component separation of full-sky maps in the radio and microwave frequencies, such as the cosmic microwave background (CMB), relies on a thorough understanding of the statistical properties of the Galactic foreground emissions. Using scalar Minkowski functionals and their tensorial generalization known as Minkowski tensors, we analyze the statistical properties of one of the major foreground components, namely the Galactic synchrotron given by the full sky 408 MHz Haslam map. We focus on understanding the nature of non-Gaussianity and statistical isotropy of the cooler regions of the map as a function of angular scale. We find that the overall level of the non-Gaussian deviations does decrease as more high emission regions are masked and as we go down to smaller scales, in agreement with the results obtained in earlier works. However, they remain significantly high, of order 3.3$\sigma$, at the smallest angular scales relevant for the Haslam map.  We carry out a detailed examination of the non-Gaussian nature using the generalized skewness and kurtosis cumulants that arise in the perturbative expansion of Minkowski functionals for weakly non-Gaussian fields.
We find that the leading sources of non-Gaussianity are the kurtosis terms which are considerably larger than the skewness terms at all angular scales. Further, for the cooler regions of the Haslam map, we find that the non-Gaussian deviations of the Minkowski functionals can be well explained by the perturbative expansion up to second-order (up to kurtosis terms), with first-order terms being sub-dominant.  
Lastly, we test the statistical isotropy of the Haslam map and find that it becomes increasingly more isotropic at smaller scales.}

\maketitle

%%%%%%%%%%%%%%%%%%%%%%%%%%%%%%%%%%%%%%%%%%%%%%%%%%%%%%%%%%%%%%%%%%%%%%%%%%%
\section{Introduction}

%Point1: Context and need for understanding statistical nature and SI. }
One of the major challenges in detecting cosmological signals such as the cosmic microwave background (CMB) radiation and 21-cm emissions from the epoch of reionization (EoR) is the removal of foreground emissions from our Galaxy and extragalactic sources. In general, these foreground components are dominant (can be several orders of magnitude higher depending on the frequency)  in comparison to the primordial signals. 
The low-frequency part of the CMB black-body spectrum is dominated by diffuse synchrotron and free-free emissions, while thermal dust emissions and the cosmic infrared background are dominant on the higher frequency side. Moreover, synchrotron and dust emissions are highly polarized. 
In many component-separation and cleaning algorithms, modelling each foreground component is a key step, and understanding both spectral and statistical properties of the foreground emissions is important in developing these pipelines~\cite{Eriksen:2004ss,Delabrouille:2008qd}.

%\textcolor{red}{Point 2: Assumption Gaussianity and SI in foreground modelling.}

  In modelling and simulating foregrounds at low frequencies, the small scale fluctuations are generally assumed to be statistically isotropic Gaussian random fields (see for example, \cite{Tegmark:1999ke,Jelic:2008jg,PySM:2017}).  
  For example, in the commonly used \texttt{Hammurabi} code~\cite{Waelkens:2008gp}, the turbulent component of the Galactic magnetic field is usually assumed to be statistically isotropic and Gaussian distributed. The assumptions of Gaussianity and statistical isotropy (henceforth, SI) of the foregrounds at small scales and away from the Galactic plane simplify their modelling. 
  However, their validity based on physical grounds is not clear.  
  
Since the interactions that govern the Galactic emissions are in general non-linear, we do not expect that the interaction may be expressed as a small perturbation term added to an interaction-free physical system. 
It is not clear that the statistical nature of each foreground component will approach Gaussianity at smaller scales. It is possible that if we remove the larger scale fluctuations, the fields do approach Gaussianity as a manifestation of the central limit theorem.  However, it is important to test this as a function of resolution or scale. Further, the foreground fields are obviously anisotropic on the full sky since most of the emissions come from regions around the plane of the Galactic equator. It is important to test whether, after masking the Galactic regions, the fields approach SI as we probe down to smaller scales.

%\textcolor{red}{Methods for Analysis}

Several techniques have been devised to probe the Gaussianity and SI of random fields in cosmology. Search methods for non-Gaussian deviations include cumulants such as skewness and kurtosis, 
bispectrum~\cite{Komatsu:2001rj}, the  Kullback-Leibler divergence~\cite{Ben-David:2015sia} etc. Some of the methods for testing SI are Bipolar Spherical Harmonics (BiPoSH) \cite{Souradeep:2006dz} and power-tensor method \cite{Rath:2015oga}. 
Morphological statistics such as scalar Minkowski functionals  (MFs)~\cite{Adler:1981,Tomita:1986} are computed in real space and  contain information of all orders of $n$-point functions. This makes them particularly advantageous over Fourier space methods such as the bispectrum and trispectrum in searches for non-Gaussianity in situations where the non-Gaussian nature of the field is a priori unknown, and/or when the field is highly non-Gaussian. 
MFs have been widely used in CMB cosmology to search for primordial non-Gaussianity~\cite{Gott:1990,
Schmalzing:1998,COBE_NG:2000,Chingangbam:2009vi,Chingangbam:2017sap,Ade:2015ava,Buchert:2017uup}. They have also been used to detect residual foreground contamination in WMAP data~\cite{Chingangbam:2013}, and to study the properties of synchrotron radiation~\cite{Rana:2018oft}. Related topological quantities like Betti numbers were also employed to understand the morphology of the interstellar turbulence \cite{2018MNRAS.475.1843M}.
Tensor-valued generalizations of the scalar MFs on two- and three-dimensional Euclidean space, which we will refer to as Minkowski tensors (MTs),
carry additional information related to intrinsic anisotropy and alignment of structures~\cite{Schroder2D:2009,Chingangbam:2017uqv}. 
The rank-2 translation invariant MTs contain the scalar MFs as their traces. 
They have been generalized to random fields on curved two-dimensional manifolds, in particular, spaces of constant curvature such as the sphere, in~\cite{Chingangbam:2017uqv}.  
MTs have been used to study departure from SI of the CMB~\cite{Vidhya:2016,Joby:2018}, to probe the effect of weak lensing on the morphology of CMB  fields~\cite{Goyal:2019vkq}, the time evolution of the fields of the EoR~\cite{Kapahtia:2017qrg,Kapahtia:2019ksk}, and matter density evolution and redshift space  distortion~\cite{Appleby:2018tzk}.

%\textcolor{red}{Aim of the paper and summary of results.}
The all-sky 408 MHz synchrotron map obtained by Haslam et al.~\cite{Haslam:1981,Haslam:1982} has been an important input for modelling the synchrotron in the CMB component separation methods for WMAP and Planck~\cite{2013ApJS..208...20B,Ade:2015qkp}. Various statistical properties that focus on the two-point function of this map have been well studied~\cite{Cho:2010kw,Mertsch:2013pua}.  The data has been de-striped and renewed by Remazeilles et al.~\cite{Remazeilles:2014mba}. 
We refer to this version as the {\em Haslam} map. 
Ben-David et al.~\cite{Ben-David:2015b} reported that this map is Gaussian at scales smaller than roughly $3^{\circ}$, using skewness and kurtosis statistics for the investigation. Rana et al.~\cite{Rana:2018oft} used the bispectrum and MFs to probe the non-Gaussianity of the Haslam map and reported findings that are in agreement with~\cite{Ben-David:2015b}.  
In this paper, we examine in detail the non-Gaussian nature and SI of the Haslam map using the Minkowski tensors as a unified statistical tool. Further, we calculate the generalized skewness and kurtosis cumulants that enter in the perturbative expansion of scalar MFs for weakly non-Gaussian fields about the zeroth-order Gaussian expressions~\cite{Matsubara:2011,Matsubara:2020}. We compare the non-Gaussian deviations of the MFs that are obtained using the analytic expressions with the exact numerical calculations. This comparison allows us to demonstrate that the perturbative expansions of the MFs about the zeroth-order forms expected for Gaussian fields are valid in the cooler regions of the Haslam map. Moreover,  the leading source of non-Gaussianity is the second-order perturbation terms, and hence, the kurtosis determines the nature of non-Gaussian deviations in the Haslam map. 

%Structure of the Paper
The paper is organized as follows. Section \ref{sec:sec2} presents a brief discussion of the Galactic synchrotron emissions, followed by a description of the Haslam map. 
In section \ref{sec:sec3}, we briefly review Minkowski tensors and scalar MFs. We also give the analytic formulae for the scalar MFs for weakly non-Gaussian random fields and present the methods we use for the numerical computation of MFs and MTs. Section \ref{sec:sec4} contains the pipeline for our analysis and simulations of Gaussian  maps with the Haslam power spectrum. Section \ref{sec:sec5} contains our calculations and main results. We end with a summary of our results and a discussion of their implications in section \ref{sec:sec6}
. Appendix \ref{sec:a1checks}  contains a discussion of the consistency checks between the Haslam map and the simulations. We show the probability distribution function (PDF) of the Haslam map in appendix \ref{sec:pdf}. 
In appendix \ref{sec:a2fnlgnl}, we show the agreement between exact numerical calculation and the analytic perturbative formulae of non-Gaussian deviations of scalar MFs for primordial local type non-Gaussianity. 

%%%%%%%%%%%%%%%%%%%%%%%%%%%%%%%%%%%%%%%%%%%%%%%%%%%%%%%%%%%%%%%%%%%%%%%%%%%
\section{Galactic synchrotron radiation and the Haslam map}
\label{sec:sec2}

Relativistic electrons interacting with magnetic fields emit synchrotron radiation. Cosmic rays (CR), which include relativistic electrons, arrive at our Galaxy from all directions. They interact with the Galactic magnetic field and emit synchrotron radiation roughly in the frequency range of 20 MHz to 100 GHz.  The intensity of the synchrotron emission depends on the number density of CR electrons and  the strength of the magnetic field, both of which vary with respect to direction. As a consequence, the intensity of Galactic synchrotron emission shows variations across the sky. 

Let the energy distribution of relativistic electrons be given by the power law form 
\begin{equation}
N_e(E) {\rm d}E \propto E^{-p}{\rm d}E,
\end{equation}
where $p$ is the index which in this case depends on the CR composition. Let $I_{\rm sync}(\hat n,\nu)$ denote the intensity of synchrotron emission in sky direction $\hat n$, at frequency $\nu$. Let $B_{\perp}$ be the magnitude of the magnetic field perpendicular to the line-of-sight radial coordinate $r$.  Then $I_{\rm sync}(\hat n,\nu)$  can be related to $B_{\perp}$ as
\begin{equation}
I_{\rm sync}(\hat n,\nu) \propto \nu^{\beta_{\textsf{s}}} \int {\d}r B_{\perp}^{-\beta_{\textsf{s}}+1}(\hat n) 
\end{equation}
The spectral index ($\beta_{s}$), which is related to $p$ as $\beta_{s}=-(p-1)/2$, shows variations in the sky given the difference in the magnetic field strength as well as the CR distribution along the line-of-sight. The spectra also exhibit steepening at higher frequency bands due to the radiative losses and the aging effects of CR electrons, and the presence of multiple components~\cite{1959ApJ...130..241W,1986rpa..book.....R,2012ApJ...747....5L}.

Our attention in this work is on $I_{\rm sync}(\hat n,\nu)$ as a fluctuating field on the sphere. Radio telescopes used in sky surveys do not measure $I_{\rm sync}$ directly. Rather, what is measured is the brightness temperature $T_{\rm sync}$ which is related to $I_{\rm sync}$ as $T_{\rm sync}=I_{\rm sync}/\nu^{2}$.
Sky surveys to obtain $T_{\rm sync}(\hat n,\nu) $ have been conducted at a number of radio frequencies. Of these, the 408 MHz Haslam map obtained by Haslam et al.~\cite{Haslam:1981,Haslam:1982} is most widely used in the CMB component separation pipelines. In this frequency range and in terms of brightness temperature, the best-fit value of spectral index ($\beta_s$) is -2.7 with an uncertainty~$\Delta\beta_{\textsf{s}}$=0.12~\cite{2003A&A...410..847P}. This low-frequency radio map is free from other interstellar radiation fields such as the free-free and spinning dust emissions which makes it an ideal synchrotron intensity map for the parametric component separation techniques.

The angular power spectrum of the Haslam map has been studied in several earlier works. It follows a power law form, $C_{\ell}=\ell^{-\alpha}$, with $\alpha$ $\sim$ 3~\cite{2008A&A...479..641L}. The angular features carry a wealth of information regarding the structure of the Galactic magnetic field.
Cho \& Lazarian~\cite{Cho:2010kw} analyzed how the angular spectrum of synchrotron emission is related to the  MHD turbulence in the interstellar medium. Lazarian \& Pogosian~\cite{2012ApJ...747....5L} carried out extensive theoretical calculations to explain the observed correlations of synchrotron fluctuations in terms of the CR electron spectra and the axisymmetric nature of the magnetic turbulence.

Despite various post-processing techniques applied, the earlier versions of the Haslam data contain residuals from strong radio sources and has artefacts such as striations due to the scanning strategies. To minimise the errors arising from these artefacts, a new cleaned 408 MHz map with fewer artefacts and reduced systematics was prepared by Remazeilles et al. \cite {Remazeilles:2014mba}. In this map, the brightness temperature values are given in kelvin unit. For our analysis in this paper, we will focus on the low brightness temperature regions of this renewed map to study its statistical properties using Minkowski tensors.

%%%%%%%%%%%%%%%%%%%%%%%%%%%%%%%%%%%%%%%%%%%%%%%%%%%%%%%%%%%%%%%%%%%%%%%%%
\section{Overview of tensorial Minkowski functionals in two dimensions}
\label{sec:sec3}

Tensorial Minkowski Functionals (also referred to as Minkowski tensors) are geometrical quantities that encode the morphological properties of structures. They are defined on flat space.  For analyzing all-sky data, such as the Haslam map, we need to analyze it on the sphere. The generalization of MTs to curved space was given in~\cite{Chingangbam:2017uqv}. We briefly outline the notations and the method described there, analytic expressions and methods for their numerical calculation.

%%%%%%%%%%%%%%%%%%%%%%%%%%%%%%%%%%%%%%%%%%%%%%%%%%%%%%%%%%%%%%
\subsection{Definition of tensorial and scalar Minkowski functionals}
%\label{sec:sec3}

In this section, we introduce tensorial and scalar Minkowski functionals in a unified way. Let us first consider a closed curve, denoted by $C$, on the unit sphere, $\S2$. Let $\da$ be the infinitesimal area element in the region enclosed by the curve, $\dl$ be the infinitesimal arc length of the curve and $\kappa$ the geodesic curvature of the curve. Let $\hat{T}$ denote the unit tangent vector to the curve.
The rank-2 Minkowski tensors (MTs) denoted by $\mathcal{W}_k$, with $k=0,1,2$, are defined to be~\cite{Chingangbam:2017uqv},
\begin{equation}
\mathcal{W}_{0}= \frac{B_0}{2}\, \mathbb{I} \int\da, \quad\
\mathcal{W}_{1}= B_1\int_{C}\,\hat{T}\otimes \hat{T} \, \dl, \quad 
\mathcal{W}_{2}= \frac{B_2}{2\pi}\,\int_{C} \hat{T}\otimes \hat{T}\, \kappa \,\dl. \label{eqn:mt}
\end{equation}
In the above, $\mathbb{I}$ is the $2\times 2$ identity matrix, and  $\otimes$ denotes the symmetric tensor product given by $\hat{T}\otimes \hat{T}=\frac{1}{2}\big(\hat{T_{i}}\hat{T_{j}}+\hat{T_{j}}\hat{T_{i}}\big)$. The coefficients $B_k$ are constants which we leave unspecified here so as to focus on the geometrical meaning of $\mathcal{W}_{k}$. 
The three scalar Minkowski functionals denoted by $V_k$ are given by the traces of $\mathcal{W}_k$, as given below,
\begin{equation}
V_{0} = B_0\,\int {\rm d}a,\quad V_{1} = B_1\,\int_C \dl,\quad V_{2} =  \frac{B_2}{2\pi}\,\int_C \kappa \, \dl.  \label{eqn:smf}
\end{equation}
 $V_0$ is proportional to the area enclosed by the curve and  $V_1$ to the perimeter of the curve. $V_{2}$, usually referred to as the {\em genus} in cosmology\footnote{It differs from the mathematical definition of the genus by one.}, equals $B_2$ for a single curve if the space is flat, while on curved space it equals $B_2$ plus a term which is proportional to $V_0$. Therefore, the Minkowski tensors combine the information contained in the scalar MFs along with new information of the shape of structures encoded in  $\mathcal{W}_{1}$.

Next, we consider smooth random fields on $\S2$. Let $u$ denote the random field. The boundaries of a level or excursion set of the field, $u=\nu$, where $\nu$ denotes the chosen field level or threshold value, form smooth closed curves. Let $Q_{\nu}$ denote the set of points in the excursion set and ${\partial Q_{\nu}}$ denote its boundary. The subscript is used to remind us that the excursion set depends on $\nu$. Then, we can generalize the definition of $\mathcal{W}_k$ to the excursion set by the following,
\begin{equation}
\mathcal{W}_{0}(\nu)= \frac{B_0}{2}\, \mathbb{I} \int_{Q_{\nu}}\da, \quad\
\mathcal{W}_{1}(\nu)= B_1\,\int_{\partial Q_{\nu}} \hat{T}\otimes \hat{T} \, \dl, \quad 
\mathcal{W}_{2}(\nu)= \frac{B_2}{2\pi}\int_{\partial Q_{\nu}}  \hat{T}\otimes \hat{T}\, \kappa \,\dl. \label{eqn:mt_field}
\end{equation}

$\mathcal{W}_{1}$, the tensorial analogue of the contour length $V_{1}$, encodes the information of the existence of any particular alignment for the structures. Structures that have no elongation in any particular direction will have $\mathcal{W}_{1}$ proportional to the identity matrix. Let 
$\overline{\mathcal{W}}_{1}$ denote the sum over the $\mathcal{W}_{1}$  for all the curves in a given threshold. Let $\Lambda_{1}$ and $\Lambda_{2}$ be its eigenvalues. Then, we can define the parameters $\alpha$ as,
\begin{equation}
\alpha=\frac{\Lambda_{1}}{\Lambda_{2}} \hspace{2cm} \Lambda_{1}<\Lambda_{2}
\end{equation}   
$\alpha$ gives the measure of the relative alignment or the deviation from SI of the field. $\alpha=1$ is obtained when $\overline{\mathcal{W}}_{1}$ is proportional to the identity matrix, and it implies that the field preserves SI, whereas deviation from unity indicates the presence of alignment for the structures.

Before we proceed, a discussion regarding our notation is in order.  
When applying to random fields in cosmology, the scalar and tensorial MFs are usually expressed per unit area in the form of densities. We use the same symbols $\mathcal W_k$ and $V_k$, with $k=0,1,2$,  to denote the densities by including the area factor in the coefficients. In the next subsection in the place of $B_k$, we will use coefficients $A_k$ which include the area factors and whose values are commonly used in the literature.  
Thus, $V_0$ gives the area fraction of the excursion set, $V_1$ the total boundary contour length per unit area, and $V_2$ the genus per unit area at each field threshold. Similarly,  $\mathcal{W}_1$, denotes the contour MT per unit area. 

%%%%%%%%%%%%%%%%%%%%%%%%%%%%%%%%%%%%%%%%%%%%%%%%%%%%%%%%%%%%%%
\subsection{Analytical formulation of scalar MFs for mildly non-Gaussian fields}
\label{sec:ana}

Let $u$ denote a generic Gaussian random field having zero mean and standard deviation $\sigma$, and let $\nu$  now denote threshold values of the normalized field $u/\sigma$. Then the expectation values of the scalar MFs per unit area, as functions of the threshold $\nu$ are given by~\cite{Tomita:1986},
\begin{equation}
V_{k}(\nu)=A_k\,e^{-\nu^{2}/2}v^{(\rm G)}_k(\nu),
\label{eqn:gmf}
\end{equation}
where $k=0,1,2$ and the coefficients $A_k$ are
\begin{equation}
A_{k}=\frac{1}{(2\pi)^{(k+1)/2}}\frac{\omega_{2}}{\omega_{2-k}\omega_{k}}\Big(\frac{\sigma_{1}}{\sqrt{2}\sigma}\Big)^{k}.
\end{equation}
The numerical factors are $\omega_0=1,\ \omega_1=2,\ \omega_2= \pi$, and $\sigma_1\equiv \sqrt{\langle |\nabla u|^2\rangle}$, where $\nabla u$ is the gradient of the field . The functions $v^{(\rm G)}_k$ are 

\begin{eqnarray}
 v_0^{\rm (G)}(\nu) &=& \sqrt{\frac{\pi}{2}}\,e^{\nu^{2}/2}\,\text{erfc}\left(\frac{\nu}{\sqrt{2}}\right),\\ 
v^{(\rm G)}_1 &=& 1,\\
v^{(\rm G)}_2(\nu) &= & \nu.
\end{eqnarray}

The superscript `G' stands for Gaussian.

For mildly non-Gaussian fields, again denoted by $u$, 
the scalar MFs can be expressed in the same form as  eqn.~\ref{eqn:gmf}, but  with  $v^{(\rm G)}_k$ replaced by $v_k$, which can be expanded in powers  of the standard deviation  $\sigma$~\cite{Matsubara:2011} as,
\begin{equation}
v_{k}=v_{k}^{(\rm G)}+v_{k}^{(1)}\sigma+v_{k}^{(2)}\sigma^{{2}}+\mathcal{O}(\sigma^{3}). 
\end{equation}  
The first-order non-Gaussian terms are given in terms of three skewness cumulants, denoted by $S_0, S_1, S_2$, as,
\begin{eqnarray}
v_{0}^{(1)}(\nu)&=&\frac{S_0}{6}H_{2}(\nu),\\
v_{1}^{(1)}(\nu)&=&\frac{S_0}{6}H_{3}(\nu)-\frac{S_1}{4}H_{1}(\nu),\\
v_{2}^{(1)}(\nu)&=&\frac{S_0}{6}H_{4}(\nu)-\frac{S_1}{2}H_{2}(\nu)-\frac{S_2}{2}H_{0}(\nu).
\end{eqnarray}
where $H_n$($\nu$) are the Hermite polynomials. The second-order non-Gaussian terms are given in terms of four kurtosis cumulants, denoted by $K_0, K_1, K_2, K_3$,  as,
\begin{eqnarray}
  v_{0}^{(2)}(\nu)&=&\frac{S_0^2}{72}H_{5}(\nu)+\frac{K_0}{24}H_{3}(\nu), \\
v_{1}^{(2)}(\nu)&=&\frac{S_0^2}{72}H_{6}(\nu) +\frac{K_0-S_0S_1}{24}H_4(\nu)
   -\frac{1}{12}  \left(K_1+\frac{3}{8}S_1^2\right)  H_2(\nu)-\frac{K_3}{8}, \\
   v_{2}^{(2)}(\nu)&=&\frac{S_0^2}{72}H_{7}(\nu)+\frac{K_0-2S_0S_1}{24}H_{5}(\nu)-\frac{1}{6}\left(K_1+\frac{1}{2}S_0S_2\right)H_{3}(\nu) \nn \\
   && -\frac{1}{2}\left(K_2+\frac{1}{2}S_1S_2\right)H_{1}(\nu). 
\end{eqnarray}
 
In terms of the field $u$, its gradient $\nabla u$, and Laplacian $\nabla^{2} u$, the skewness and kurtosis cumulants are defined as follows
\begin{eqnarray}
S_0 &=& \frac{\langle u^{3}\rangle_c}{\sigma^{4}}, \quad S_1=\frac{\langle u^{2}\nabla^{2}u\rangle_c}{\sigma^{2}\sigma^{2}_{1}},  \quad S_2= \frac{2\langle |\nabla u|^{2}\nabla^{2}u\rangle_c}{\sigma_{1}^{4}}, \label{eqn:skew}\\
K_0 &=& \frac{\langle u^{4}\rangle_c}{\sigma^{6}}, \ K_1=\frac{\langle u^{3}\nabla^{2}u\rangle_c}{\sigma^{4}\sigma^{2}_{1}}, \  
K_2=\frac{2\langle u|\nabla u|^{2}\nabla^{2}u\rangle_c+\langle |\nabla u|^{4}\rangle_c}{\sigma^{2}\sigma_{1}^{4}}, \ K_3=\frac{\langle |\nabla u|^{4}\rangle_c}{2\sigma^{2}\sigma_{1}^{4}}. \label{eqn:kurt}
\end{eqnarray}
The subscript $c$ indicates that these quantities are the connected cumulants. As the field is mean-free, the third-order cumulants are equal to the third-order moments. Fourth-order cumulants are given in terms of the moments~\cite{Matsubara:2020}  as,
\begin{eqnarray}
\langle u^{4} \rangle_{c}&=& \langle u^{4} \rangle -3\sigma^{4} \\
\langle u^{3}\nabla^{2}u\rangle_c &=& -3\langle u^{2}|\nabla u|^{2} \rangle_{c}=-3(\langle u^{2}|\nabla u|^{2} \rangle -\sigma^{2}\sigma_{1}^{2}) \\
\langle u|\nabla u|^{2} \nabla^{2} u\rangle_{c} &=& \langle u|\nabla u|^{2} \nabla^{2} u\rangle +\sigma_{1}^{4} \\
\langle |\nabla u|^{4} \rangle_{c} &=& \langle |\nabla u|^{4} \rangle -2\sigma_{1}^{4} 
\end{eqnarray}

%%%%%%%%%%%%%%%%%%%%%%%%%%%%%%%%%%%%%%%%%%%%%%%%%%%%%%%%%%%%%%%%%%%%%%
\subsection{Computing scalar and tensorial Minkowski functionals} 

\subsubsection{Method 1 - using field derivatives }
\label{sec:method1}

Given the field $u$, we rescale it to make it mean-free and unit standard deviation. MFs can be computed at each threshold values $\nu$ in terms of $u$ and its derivatives. The line integrals in eq.~\ref{eqn:mt_field} can be converted into surface integrals. $V_0$ is expressed in terms of the field $u$ as,
\begin{equation}
\text{V}_{0}(\nu)=\int_{S^{2}}\Theta(u-\nu)\hspace{1mm}{\rm d}a,
\end{equation}
where $\Theta$ is the step function. 
Using $\hat{T}_{i}=\epsilon_{ij}\frac{u_{;j}}{|\nabla u|}$, 
where $\epsilon$ is the two dimensional anti-symmetric Levi-Civita tensor and $u_{;j}$ the $j$-th component of the covariant derivative, we get 
\begin{eqnarray}
\overline{\mathcal{W}}_{1}&=&\frac{1}{4}\int_{S^{2}} \delta  (u-\nu)\hspace{1mm}\frac{1}{|\nabla u|}\hspace{1mm}\mathcal{M} \hspace{1mm}\da,\\
\overline{\mathcal{W}}_{2} &=& \frac{1}{2\pi}\int_{S^{2}} \delta  (u-\nu)\hspace{1mm}\frac{\kappa}{|\nabla u|}\hspace{1mm}\mathcal{M} \hspace{1mm} \da.
\end{eqnarray}
The expression for $\kappa$ is
\begin{equation}
\kappa=\frac{2u_{;1}u_{;2}u_{;12}-u_{;1}^{2}u_{;22}-u_{;2}^{2}u_{;11}}{|\nabla u|^{3}}
\end{equation}
and $\mathcal{M}$ is,
	\[
	\mathcal{M}=
	\begin{bmatrix}
	u_{;2}^{2} & -u_{;1}u_{;2}  \\
	-u_{;1}u_{;2} & u_{;1}^{2}  
	\end{bmatrix}
	\]
For a field in discretized space, the $\delta$-function is approximated to be
$\delta(u-\nu)=\frac{1}{\Delta \nu}$, 
if $u$ lies between $\nu-\Delta \nu/2$ and $\nu+\Delta \nu/2$, and zero elsewhere. Here, $\Delta \nu$ is the bin size of the threshold values. 
Using this method, we compute $V_0$, $\mathcal{W}_1$ and  $\mathcal{W}_2$. From the eigenvalues of $\mathcal{W}_1$, we calculate  $\alpha$. Further, by taking the traces of  $\overline{\mathcal{W}}_{1}$ and $\overline{\mathcal{W}}_{2}$, we obtain $V_1$ and $V_2$. All these quantities are then divided by the total area to get their corresponding densities. 
The $\delta$ function approximation is shown to have inherent numerical error in~\cite{Lim:2012}. This error will be present in the calculations of  $\mathcal{W}_1$ and  $\mathcal{W}_2$,  and in $V_1$ and $V_2$. For comparison, we will compute $V_1$ and $V_2$ using the geometric method, which is described in the next subsection.  It was shown in~\cite{Goyal:2019vkq} that the numerical errors in the two eigenvalues of $\mathcal{W}_1$ are comparable and, hence, get cancelled out when computing $\alpha$. Therefore, the calculation of $\alpha$ is unbiased. 

%%%%%%%%%%%%%%%%%%%%%%%%%%%%%%%%%%%%%%%%%%%%%%%%%%%%%%%%%%%%%%%%%%%%%
\subsubsection{Method 2 - geometric method for scalar MF estimation }
\label{sec:method2}

The geometric estimation of MFs is carried out by first identifying the structures from the excursion sets at different field thresholds. 
 In the following, we briefly outline the method followed by the \texttt{CND$\_$REG2D}~\cite{Ducout:2013} code that we have used, and refer the reader to the original papers for the details. 
 
The excursion sets are obtained as a binary field by identifying pixels where the fields have values below or above the chosen threshold $\nu$. The algorithm identifies each structure (connected region or hole) by marking boundary pixels as different from the pixels in the inner part of the structures. 
The total number of pixels included in each structure will constitute the first MF, area fraction ${V}_{0}$.  Next, the 
genus, ${V}_{2}$, is obtained by computing the vertices of the pixel grid at the boundaries of the structures. It uses the Gauss-Bonnet theorem, which relates the genus to the integration of the curvature along the boundary. As topological properties are invariant under the continuous transformation of the boundary, summing the vertices with appropriate weights will give the genus of the region. The estimation of the next MF, the contour length, ${V}_{1}$, is a bit more involved. The length of the perimeter of the polygon formed by the boundary gives ${V}_{1}$. Due to the pixellated form, the error in the estimation of the length can be large. In order to control the error, interpolation of field values between adjoining boundary pixels is performed so as to smoothen the pixelated boundary and obtain a sufficiently accurate estimate of the length of the polygon.

%%%%%%%%%%%%%%%%%%%%%%%%%%%%%%%%%%%%%%%%%%%%%%%%%%%%%%%%%%%%%%%%%%%%%%%%%%%
\section{Analysis pipeline and Gaussian isotropic simulations}
\label{sec:sec4}

We focus our analysis on cooler parts of the sky by applying different brightness temperature cuts to mask the regions above the chosen cutoff temperatures. We also analyze at different angular scales. 
For this purpose, we process the map following an appropriate pipeline. For comparison, we also generate 1000 Gaussian isotropic simulations using the Haslam power spectrum. The pre-processing pipeline and the Gaussian isotropic simulations are described below.

%%%%%%%%%%%%%%%%%%%%%%%%%%%%%%%%%%%%%%%%%%%%%%%%%%%%%%%%
%\section{Pipeline for pre-processing the maps}
%\label{sec:pipeline}
%Below we outline the steps that we follow for pre-processing the maps. 

%%%%%%%%%%%%%%%%%%%%%%%%%%%%%%%%%%%%
\subsection{Masking} 
\label{sec:mask}
Since our target region for analysis are the regions away from the Galactic center where the emission is very strong, we mask the map as described below. 
\begin{itemize}
	\item First, we degrade the Haslam map from $N_{\rm side}$= 512 to 128
since the resolution of processed 408 MHz map is $56^{\prime}$~\cite{Remazeilles:2014mba}. This map has mean value 34.4~K. We rescale each pixel value with the mean value. Then, we apply a Galactic cut ($ \mid b\mid< 10^{\circ}$) to avoid the 
contamination from the Galactic plane.
	\item We mask the bright Loop-I ring, coming from an old supernovae remnant. The mask region includes $\pm 4^{\circ}$ width cut around a
circle of radius $58^{\circ}$ entered at ($l$,$b$) = ($329^{\circ}$; $17^{\circ}$. 5).
	\item Let us denote the brightness temperature cut by $u_c$. We construct five sky masks by applying different choices of $u_c$. Pixels having field values above $u_c$ are set to zero. We carry out the analysis for the values $u_c=22,25, 30, 40, 60$ K. Our results will be presented for the cases $u_c=25$ and 60 K. The effective fractional sky coverages are  0.74 for  $u_c= 60$ K,  0.69 for $u_c= 40$ K,   0.56 for $u_c= 30$ K, 0.41 for $u_c= 25$ K and 0.27 for $u_c= 22$ K .
	\end{itemize}
To avoid the leakage of power due to sharp cutoff between the
masked and the unmasked regions in the map, we apodize the masks by convolving with a $5^{\circ}$ FWHM Gaussian. Such a smooth Gaussian filter function minimizes the leakage of the signal towards the edges of the unmasked region.
%%%%%%%%%%%%%%%%%%%%%%%%%%%%%%%%%%%%
\begin{figure}[t]
\centering
 \includegraphics[width=8.1cm,height=4.7cm]{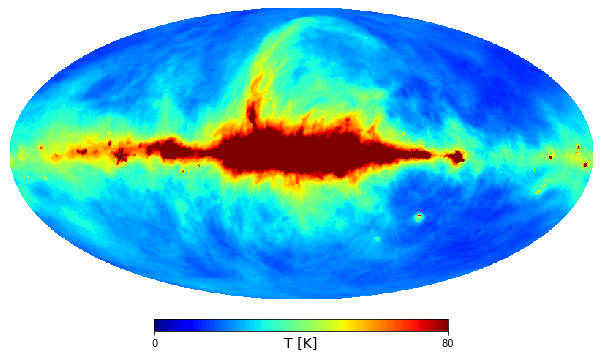}\quad\quad
  \includegraphics[scale=0.36]{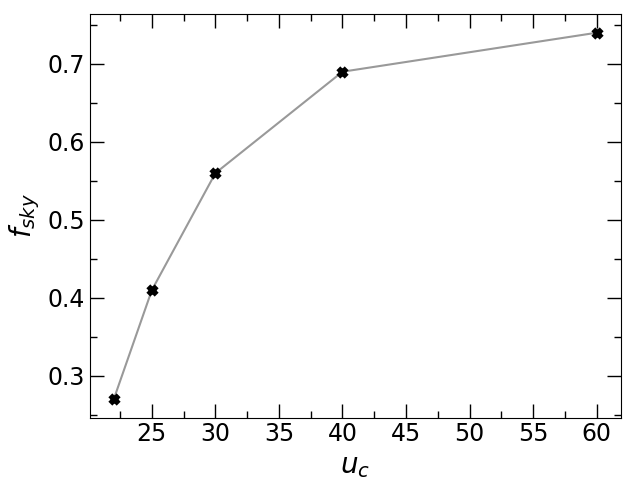}
 \
 \\
 \vskip .2cm
 \includegraphics[width=8.9cm,height=4.8cm]{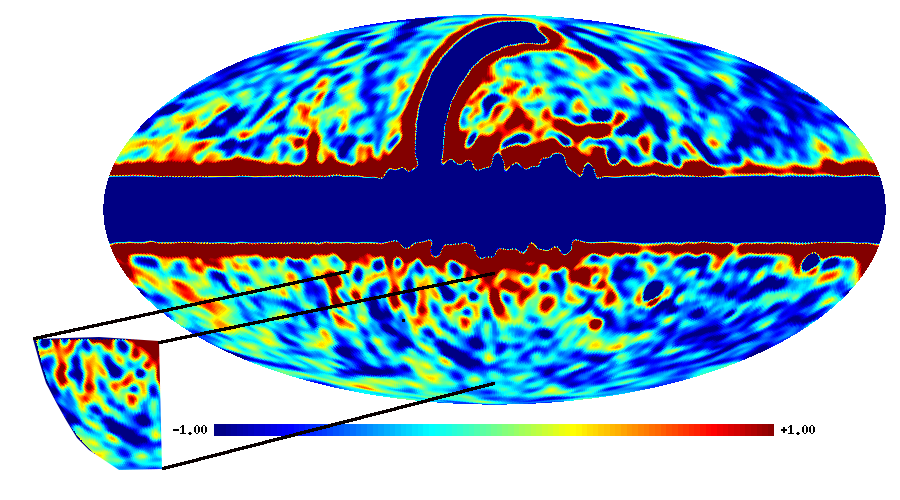}\quad\quad\quad
 \includegraphics[scale=0.6]{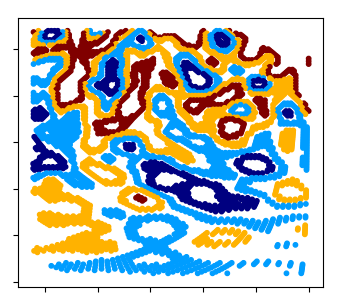} 
  \caption{{\em Top}: The left panel shows the reprocessed version of the all sky 408 MHz Haslam map. The field values are given in kelvin unit and have mean value 34.4~K. The right panel shows the sky-fraction ($f_{sky}$) as a function of different temperature cuts ($u_c$) used in our analysis.
    {\em Bottom}: The left panel is the mean-free and normalized ($u/\sigma$) version of the same map after applying the brightness temperature cut $u_c=60$\,K and band-passing with multipole cut $\ell_{c}=30$. 
    Excursion set boundaries are shown in the right panel for different field thresholds (different colours), corresponding to the cut-out patch on the lower left of the left panel. The boundaries are quite thick due to the choice of large field threshold bins.}
\label{fig:fig1}
\end{figure}

\subsection{Band-passing}
\label{sec:bp}

In order to focus our analysis on specific angular scales of interest, we use a band-pass filter of the following form: 
\begin{equation}
f(\ell)=\frac{1}{4}\left\{1+ {\rm tanh}\left(\frac{\ell-\ell_{c}}{\Delta \ell}\right)\right\}\left\{1-\text{tanh}\left(\frac{\ell-180}{\Delta \ell}\right)\right\}.
\end{equation}
This filter cuts off the Fourier amplitudes ($a_{\ell m}$s) below a multipole scale $\ell_c$,  and above $\ell=180$. The upper multipole cutoff is in accordance with the $56'$ beam size of the Haslam map. $\Delta\ell$ sets the width of the cutoff region of the filter. We use  $\Delta\ell = 10$ for the results presented in this paper. We have checked that the results are robust for a reasonable range of $\Delta\ell$.  
We vary $\ell_c$ to study the statistics of the map at different scales. 
We have also used other suitable filters such as a cosine filter and our results are found to be robust.

The final maps, which have undergone the aforementioned pre-processing steps, are, then, mean-subtracted and rescaled with the standard deviation. Computations of tensorial and scalar Minkowski functionals are done on these mean-free, unit standard deviation versions of the Haslam map.

The top left panel of figure~\ref{fig:fig1} shows the Haslam map. The top right panel gives the sky-fraction ($f_{sky}$) left for the analysis after the application of various temperature cuts ($u_c$) on the map.  The bottom left panel shows the mean-free and normalized ($u/\sigma$) version of the same map obtained after applying temperature cut  $u_c=60$\,K, and band-passed with multipole cut $\ell_c=30$. The bottom right panel shows iso-field contours of a slice of the field shown in the lower left of the left panel. Different colours of the contours correspond to different field thresholds ($\nu$). The lines are thick due to the choice of large field threshold bins.
%%%%%%%%%%%%%%%%%%%%%%%%%%%%%%%%%%%%%%%%%%%%%%%%%%%%%%%%%%%%%%%%%%%%%%%
\subsection{Gaussian isotropic simulations}

In order to quantify the non-Gaussianity and anisotropy of the Haslam map, we need to compare it with suitable Gaussian isotropic simulations. For this purpose, we obtain 1000 Gaussian isotropic simulations that are generated by using the full power spectrum of the Haslam map corrected for cut-sky, pixel and beam correction. This input spectrum is made using the publicly available code \texttt{PolSpice} \cite{Polspice,Chon:2004}. A detailed consistency check carried out for these simulations is discussed in  appendix~\ref{sec:a1checks}. The pre-processing pipeline discussed in sections \ref{sec:mask} and \ref{sec:bp} is applied to each simulation map so that both simulation and data maps are identically pre-processed. This is necessary to ensure that the comparison of the statistics that we compute from the data and simulation maps makes sense.
%%%%%%%%%%%%%%%%%%%%%%%%%%%%%%%%%%%%%%%%%%%%%%%%%%%%%%%%%%%%%%%%%%%%%%%%%% 

\section{Analysis of Gaussianity and SI of Galactic synchrotron emission -- results}
\label{sec:sec5}

This section presents the results obtained from analyzing the Haslam map and its comparison with Gaussian isotropic simulations. 
%For all calculations and comparisons, the Haslam map and Gaussian simulations are identically processed 
%by carrying out the masking and bandpassing procedures described in section~\ref{sec:pipeline}. 
%%%%%%%%%%%%%%%%%%%%%%%%%%%%%%%%%%%%%%%%%%%%%%%%%%%%%%%%
\subsection{Spectra  of the Haslam map}

\begin{figure}[t]
\begin{center}
\includegraphics[scale=0.55]{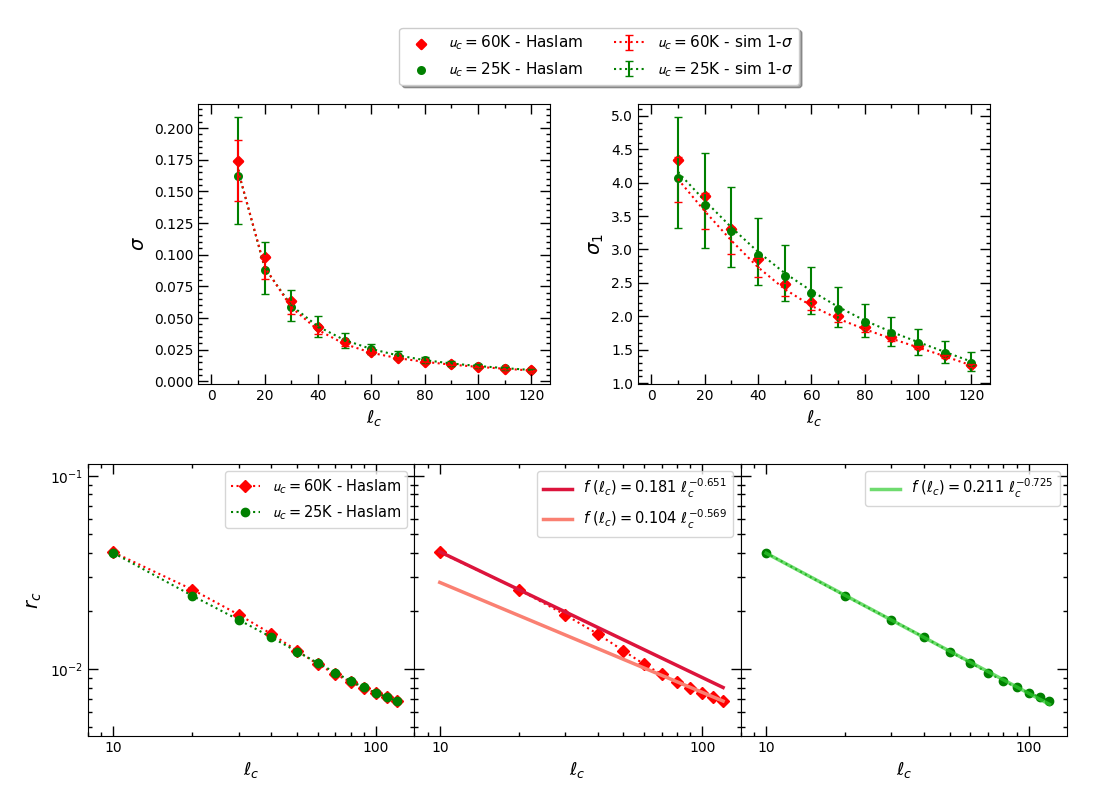}
\end{center}
\caption{{\em Top}: $\sigma$ (left) and $\sigma_1$ (right) of the Haslam map for $u_c=60$ K (red diamond) and 25 K (green circle) as a function of $\ell_c$. Mean and 1$\sigma$ error bars obtained from 1000 Gaussian isotropic simulations are also shown.  Since the simulations are obtained using the power spectrum of the Haslam data,  $\sigma$ and $\sigma_1$ for the observed data and simulations match within $1\sigma$, as expected. 
{\em Bottom}: The left panel shows the correlation length, $r_c\equiv \sigma/\sigma_1$ versus $\ell_c$ for $u_c=60$ K and 25 K. The middle panel shows $r_c$ for $u_c=60$ K fitted with two different
power law functions towards the low and high $\ell_{c}$ regimes, indicating a transition in the nature of the field at the intermediate $\ell_{c}$ scales. The right panel shows  $r_{c}$ for $u_c=25$ K fitted by a single function.}
\label{fig:rc}
\end{figure}

We first discuss the spectral parameters $\sigma$ and $\sigma_{1}$, and their ratio $r_{c}\equiv \sigma \slash \sigma_{1}$. Given the mean-free field $u$, $\sigma\equiv\sqrt{\langle u^{2}\rangle}$ and $\sigma_1\equiv\sqrt{\langle |\nabla u|^{2}\rangle}$. 
The top panels  of figure~\ref{fig:rc}, show $\sigma$ (left) and $\sigma_1$ (right) for varying $\ell_{c}$, for the Haslam map for different values of $u_c$. The mean over 1000 Gaussian simulations is also shown along with the 1$\sigma$ error bars.  
We can see that both $\sigma$ and $\sigma_1$ decrease with decrease of $u_{c}$ and towards higher $\ell_{c}$, indicating a drop in the level of fluctuations of the field and its gradient, as we go to lower temperatures as well as smaller scales. Moreover, both these parameters for the Haslam map fall within 1$\sigma$ error bars obtained from Gaussian simulations. This is expected as the simulations are generated from the power spectrum of the Haslam map and validates the correctness of these simulations.

The bottom panels show $r_c$. This quantity gives a measure of the typical size of structures (hot and cold spots) in the field. The left panels shows $r_c$ for both $u_c=60$ and 25 K, so as to enable their visual comparison. We see that higher $u_c$ has slightly larger structures towards lower $\ell_c$. 
This indicates that if we include sky regions with higher temperature, then there are  larger regions having correlated temperature values. $r_c$ also decreases with increasing $\ell_c$, which is expected due to the subtraction of large scale fluctuations, and the fact that $\sigma$ decreases faster than $\sigma_1$. We have fitted the fall of $r_c$ with respect to $\ell_c$ with power law functions (shown in the middle and right bottom panels of figure~\ref{fig:rc}). For $u_c=60 $ K, we could fit it with two functions towards the low and high $\ell_c$ regimes, indicating a transition in the nature of the field at the intermediate scales. For $u_c=25$, we are able to fit $r_c$ with a single function. This could be a hint to the difference in the nature of the field in the cooler regions of the synchrotron sky (smaller $u_c$), as seen in our further analysis.

%%%%%%%%%%%%%%%%%%%%%%%%%%%%%%%%%%%%%%%%%%%
\subsection{Skewness and Kurtosis of the Haslam map}
\label{sec:cumulants}
\begin{figure}[t]
%\begin{center}
\includegraphics[scale=0.68]{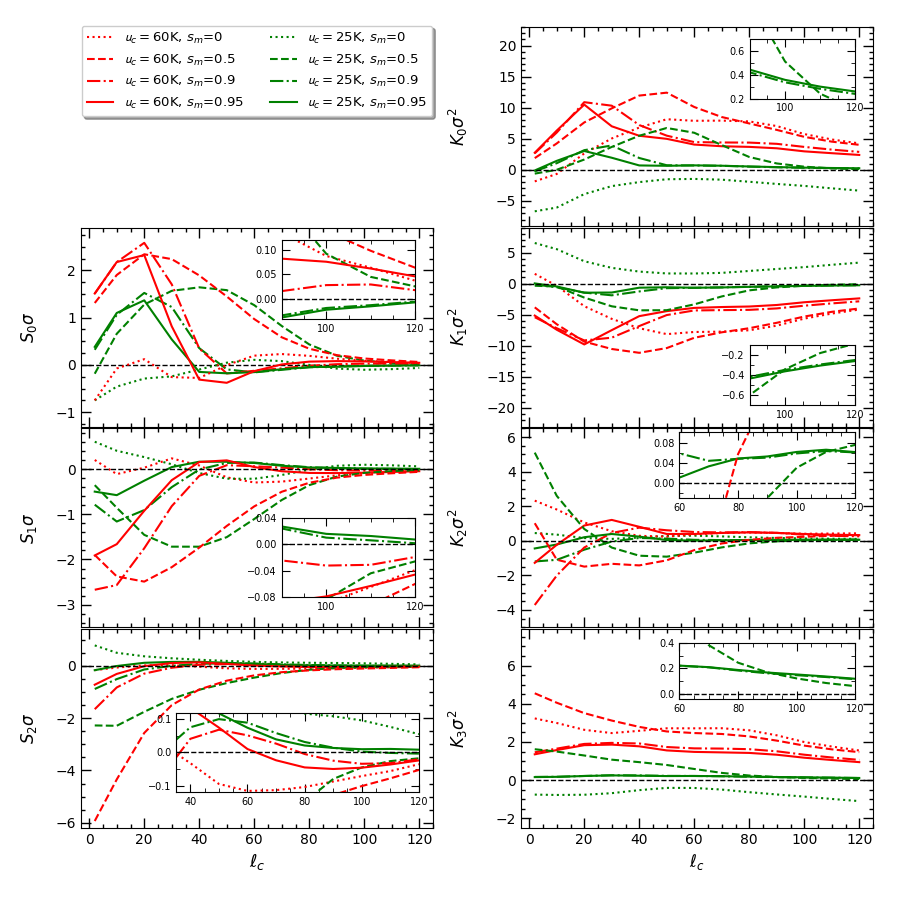}
%\end{center}
\caption{\small Skewness and kurtosis cumulants (defined in section \ref{sec:ana}) of the Haslam map for the case of temperature cut values $u_{c}=60\,$K (red) and $u_{c}=25\,$K (green), plotted as functions of the multipole cut $\ell_c$. Different line types for each colour represent different values of $s_m$, and the associated sky fraction.}
\label{fig:moments}
\end{figure}

In order to probe the non-Gaussian nature of the Haslam map, we next analyze the skewness, $S_i$, and kurtosis cumulants, $K_i$, defined in eqs.~\ref{eqn:skew} and \ref{eqn:kurt}. Since $\sigma$ varies with $\ell_c$, we interpret the cumulants with the appropriate power of $\sigma$ multiplied to them, i.e., $S_i\sigma$ and $K_i\sigma^2$. 
These quantities are the coefficients of the Hermite polynomials in the expressions for the first and second-order non-Gaussian deviations of the MFs, up to numerical factors. Hence, it is meaningful to compare them directly.

To minimize the effects of the mask boundary on the calculation of $S_i\sigma$ and $K_i\sigma^2$, and the scalar and tensorial MFs, we must stay sufficiently far away from the boundary. Upon smoothing the binary mask, pixels near the mask boundaries acquire values between zero and one. A rough estimate shows that a smoothed mask pixel value $>0.89$ roughly corresponds to $>2\theta_s$ distance from the mask boundary,  $\theta_s$ being the smoothing scale.  We introduce a parameter, $s_m$, whose value is between zero and one, to control how far away a smoothed mask pixel is from the boundary. Pixels for which the smoothed mask has values $> s_m$ are included in the calculations. As $s_m$ increases towards one, the sky fraction will decrease, and hence, the statistical significance of the results will decrease. Therefore, it is best to select an optimum value of $s_m$ such that the numerical error is minimized and the statistical significance is maximized. 

Figure~\ref{fig:moments} shows $S_i\sigma$ (left column) and $K_i\sigma^2$ (right column) as functions of the multipole cut $\ell_c$,  for $u_c=60$\,K (red lines) and 25\,K (green lines). We show the results for different values of $s_m$. The inset boxes show the zoomed in plots towards higher $\ell_c$ values. For lower $\ell_c$, we see a large variation of the cumulants with $s_m$. Towards higher $\ell_c$, they show approximately convergent behaviour for the larger $s_m$ values, indicating that the effect of the mask boundary is minimized. Therefore, we will interpret the non-Gaussian behaviour of the Haslam map using $s_m= 0.9$, for which the cumulants are shown by dot-dash lines in the plots.. 
The results for the cumulants are as follows:
\begin{itemize}
\item All four kurtosis cumulants have values whose magnitudes are considerably larger than those of the skewness ones, for both values of $u_c$ and for all $\ell_c$. This corroborates our inference from visual inspection of the probablity distribution functions (PDFs) of the Haslam map shown in figure~\ref{fig:pdf} in appendix \ref{sec:pdf}. 

\item $S_i\sigma$  show clear decrease both with decreasing $u_c$ and towards high $\ell_c$. We see that $S_0\sigma$ and $S_1\sigma$  show rough oscillatory behaviour up to intermediate values of $\ell_c$. 
 Towards high $\ell_c$, all three $S_i\sigma$  decrease monotonically.    %

\item All the kurtosis cumulants also decrease from higher to lower values of $u_c$, and at all $\ell_c$.  
Towards high $\ell_c$, the magnitudes of all the $K_i\sigma^2$ show mild monotonic decrease, except $K_2\sigma^2$ which appears to saturate at a small but finite value. $K_0\sigma^{2}$  and $K_1\sigma^{2}$ also exhibit rough oscillatory behavior similar to skewness parameters. 
  
\end{itemize}

Based on the above points, we conclude that the non-Gaussian nature of the Haslam map at the range of scales probed here is predominantly sourced by kurtosis terms. Moreover, at smaller scales, the field shows convergence towards Gaussian behaviour. However, we note that it is important to probe down to even smaller scales, which is not feasible with the Haslam map. The next subsection will discuss the level of non-Gaussianity measured using the Minkowski functionals. 
Although beyond the scope of this paper, from the behaviour of the skewness and kurtosis parameters as functions of $\ell_c$, it will be interesting to investigate further whether one can identify physically interesting scales associated with the distribution of cosmic rays and free electrons, and the properties of the Galactic magnetic field.

%%%%%%%%%%%%%%%%%%%%%%%%%%%%%%%%%%%%%%%%%%%%%%%%%%%%%%%%%%%%%%	
\subsection{Scalar Minkowski functionals for the Haslam map}
\label{sec:mt}
\begin{figure}
\includegraphics[scale=0.69]{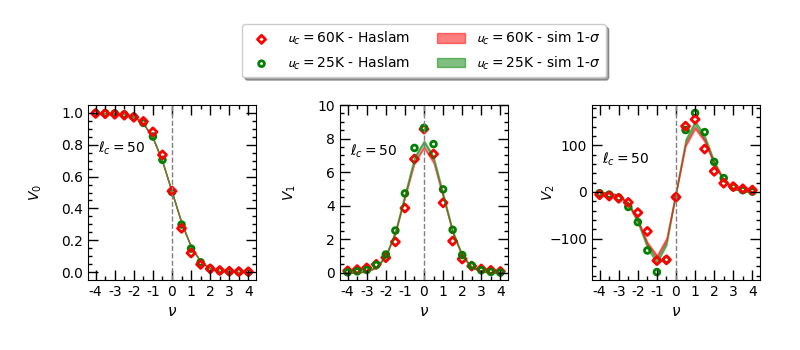}

\caption{\small Scalar MFs of the Haslam map for $u_c=60$ K (red diamonds) and $u_c=25$ K (green circles), for $\ell_{c}=50$. The ensemble mean and 1$\sigma$ width obtained from 1000 simulated maps are also plotted to show the deviation.} 
\label{fig:scalarsMFs}
\end{figure}

% Describe exact numerical results
We compute the scalar MFs  using methods 1 and 2 described in sections~\ref{sec:method1} and \ref{sec:method2}, for the  Haslam and the simulated Gaussian isotropic maps. We refer to the results of these calculations as `exact numerical results'. We use the threshold range $-4\le \nu\le 4$ with a bin size of $\Delta\nu=0.5$. We show our results for the mask boundary threshold value $s_m=0.9$, as discussed in section~\ref{sec:cumulants}. 
The results for $u_c=60$ K and 25 K, for the intermediate scale $\ell_c=50$ are shown in figure~\ref{fig:scalarsMFs}. The deviations of $V_k^{\rm Haslam}$ from the Gaussian mean values are easily discerned by eye.

In order to quantify the difference of the MFs between the Haslam map and the Gaussian  simulations, we define 
\begin{equation}
  \Delta V_k \equiv V_k^{\rm Haslam} - V^{\rm (G)}_k,
  \end{equation}
where the superscript `Haslam' refers to the Haslam map and `G' stands for Gaussian simulation. At each $k$, we compute $\Delta V_k$,  normalized by the amplitude of $V_k^{\rm (G)}(\nu)$ (indicated by superscript `max'), for each Gaussian isotropic realization.

\begin{figure}
\includegraphics[height=7.2in,width=6.5in]{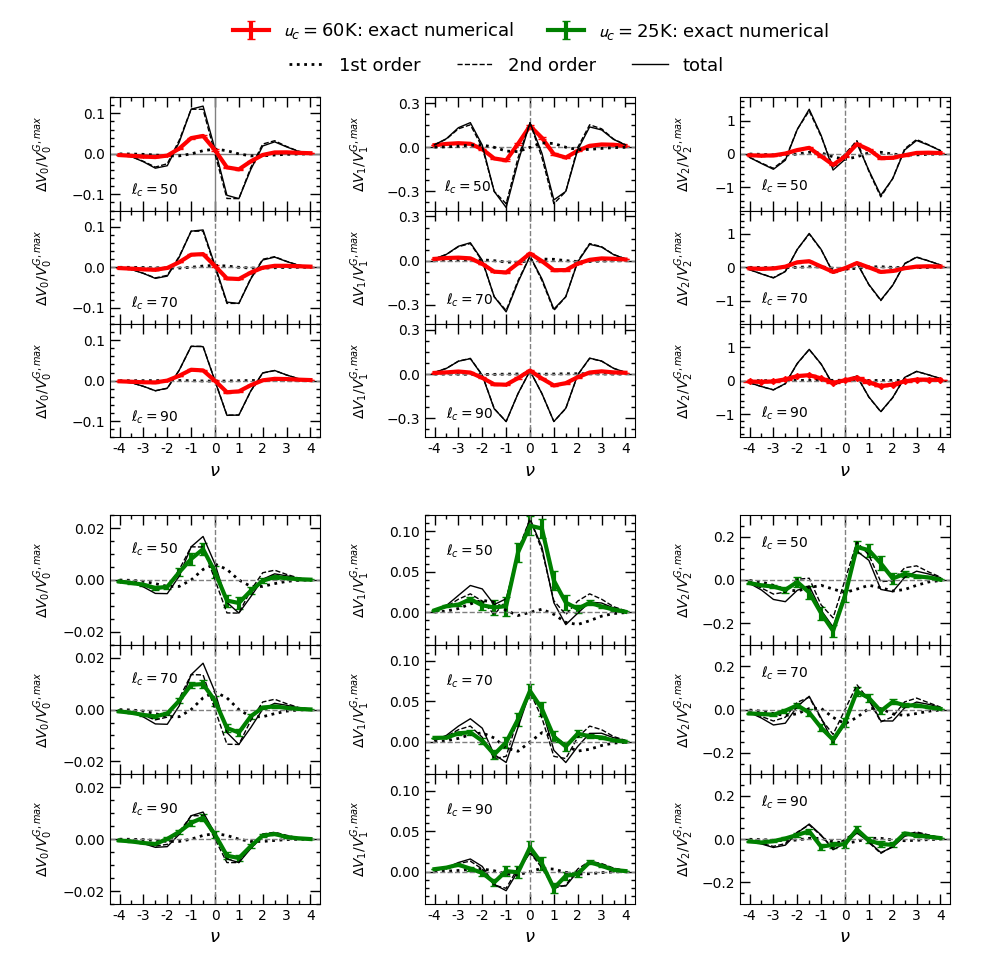}
\caption{\small  The deviations, $\Delta V_{k}/V^{\rm {G,max}}_k$, of the three MFs for Haslam data from Gaussian expectation for $u_c=60$\,K (red solid lines) and $u_c=25$\,K (green solid lines) are shown for $\ell_{c}=50,70,90$. We use $s_m =0.9$ for these results. The black lines correspond to the results obtained using perturbative expansion of MFs with only first-order terms  (dotted lines), only second-order terms (dashed lines), and the sum of first- and second-order (solid lines). The dashed and solid black lines almost overlap since the contributions from the first-order terms are small.} 
\label{fig:scalars}
\end{figure}

Figure~\ref{fig:scalars} shows  the ensemble mean and 1$\sigma$ error bars of the normalized $\Delta V_k$  for $u_{c}=60$ K (red solid lines) and $u_{c}=25$ K (green solid lines), and for $\ell_{c}=50, 70, 90$ (top to bottom rows). For $u_c=60$ K, the error bars are relatively smaller and, hence, hard to see by eye. All the plots shown in figure~\ref{fig:scalars} are obtained using method 1. We obtain almost identical results from calculations using method 2. 
We summarize our findings from the exact numerical calculations as follows:
\begin{itemize}
\item The deviations have a characteristic shape as functions of the threshold. The overall amplitude of the deviations decreases as $u_c$ is decreased, while the shape is approximately maintained. This indicates that masking out very high-intensity regions, which correspond to values on the positive tail of the PDF of the field, makes it tend towards Gaussian nature. The nature of the non-Gaussian deviation of the field approximately remains the same even as we mask more high-temperature regions (decreasing $u_c$), and as we probe down to smaller angular scales (increasing $\ell_c$). 

\item As we increase $\ell_c$, the number of structures (equivalently, fluctuations of the field per unit area on the sphere) increases. As a consequence, the error bars on $\Delta V_{k}$ decrease with increasing $\ell_c$. Therefore, even though the amplitude of the deviations decrease with increasing $\ell_c$, the statistical significance of the deviations does not decrease proportionately and can remain high. This is particularly evident for the case of $u_c=60$. At each $\ell_c$ and each $\nu$, the error bar for $u_c=25$ is generally higher than $60$ because of lower sky fraction. 
\item Lastly, we compare $\Delta V_k$ obtained from method 1 and 2. As mentioned earlier, method 1 contains small numerical inaccuracies due to the discretization of the $\delta$ function. However, when subtracting between the MFs obtained from the Haslam map and the Gaussian simulations, these numerical errors will mostly cancel out. A small part can still remain because the Haslam map is non-Gaussian, particularly for higher $u_c$ and lower $\ell_c$ values. In comparison, method 2 is free of these errors. We obtain very similar results for $\Delta V_k$ from method 2, indicating that the residual numerical errors for method 1 are insignificant and can be ignored. 
  Another important point to mention is that we obtain marginally higher error bars from method 1, compared to method 2. The reason for this is the shot noise arising from the discrete harmonic transform associated with calculating field derivatives. 
\end{itemize}

% Describe perturbative results and comparison with exact numerical results
Next, we discuss the results obtained using the perturbative formulae of MFs given in section \ref{sec:ana}, and compare with the exact numerical results.  
To do so, we define
\begin{eqnarray}
  \Delta V_k^{(1),\rm  pert} &=& A_k v_k^{(1)}, \\
\Delta V_k^{(2),\rm  pert} &=& A_k\,v_k^{(2)}, \\
\Delta V_k^{\rm total, pert} &=& A_k\,\left( v_k^{(1)} + v_k^{(2)}\right),
\end{eqnarray}
where $k=0,1,2$, and $A_k$ is the amplitude of the analytic expressions for MFs for the Gaussian case. The superscript `pert' refers to the perturbative expansion described in section~\ref{sec:ana}. For consistency with the exact numerical case, we normalize them by $V_k^{\rm{G,max}}$ (which is not always the same as $A_k$). 
In figure~\ref{fig:scalars}, the first-order results are shown by dotted black lines, second-order by dashed black lines, and the total by solid black lines. 
First, for $u_c=60$\,K (top panels accompanying the red plots), we can see that  
the first-order deviations for which skewness cumulants contribute are much smaller than the second-order deviations for which kurtosis cumulants  contribute. As a consequence, the plots for the second-order and total deviations nearly overlap. 
Secondly, the total deviations upto second-order  overestimate the amplitude of the deviations of all three MFs by over a factor of two, but the shape roughly agrees (as indicated by the location of zeros, peaks and troughs). This remains so even at high $\ell_c$. Therefore, we conclude that for $u_c=60$\,K, the Haslam field is highly non-Gaussian even at the smallest scales probed here. Hence, it is not meaningful to consider it as a Gaussian field plus a small non-Gaussian component.

In the lower panels of figure~\ref{fig:scalars}, we show the case of $u_c=25$\ K. We again 
find that  
the first-order deviations are smaller than the second-order for all MFs and all $\ell_c$ values considered here. So the main contribution to the non-Gaussian behaviour of the relatively cooler signals of the Haslam map comes from the four kurtosis cumulants. Secondly, we find very good agreement between the  deviations of the MFs given by the analytic and exact numerical calculations. The amplitude of the non-Gaussian deviations decreases as $\ell_c$ increases, indicating that the field approaches Gaussian behaviour at smaller scales.
Therefore, we conclude that the cooler regions of the Haslam map can be well approximated as a mildly non-Gaussian field. The nature of the mild non-Gaussianity, however, does not significantly vary with angular scale, as implied by the shape of the deviations of the MFs. Our analysis with other temperature cuts for the cooler regions, such as $u_c=22$ K and $u_c=30$ K, shows trends similar to what is observed with $u_c=25$ K. We are planning to do a detailed study on these regions, in our future works, to understand more on the nature of synchrotron non-Gaussianity. 

A visual comparison of  figure~\ref{fig:scalars} with the plots in figure~\ref{fig:comparefnlgnl} shows that the non-Gaussian nature of the Haslam map is similar to the local type $g_{\rm NL}$ non-Gaussianity of primordial inflationary fluctuations. This indicates the presence of an approximate parity symmetry in the fluctuations of synchrotron radiation.

%%%%%%%%%%%%%%%%%%%%%%%%%%%%%%%%%%%%%%%%%%%%%%%%%%%%%%%%%%%%%%%%%%%%%%
\subsubsection{Quantifying the level of non-Gaussianity}
\label{sec:D}

To quantify the statistical significance of the non-Gaussian deviations for $V_k$, we compute the difference between each statistic computed for Haslam data and its mean value obtained from Gaussian isotropic simulations in units of the standard deviation. This is encapsulated by $\chi^{2}$ which is defined, at each threshold $\nu$, as follows, 
\begin{equation}
\chi^2_{V_k}(\nu) = \frac{\left(V_k^{\rm Haslam}(\nu)-\overline{V}_k^{\rm (G)}(\nu)\right)^{2}}{\sigma^{2}_{V_k^{\rm (G)}}(\nu)}
\end{equation}
We compute $\chi^2_{V_k}$ for all the values of $u_c$ and $\ell_c$ that we consider here. We will not show $\chi^2$ for $u_c=60$ K since there is a high level of non-Gaussianity, and it is not meaningful to compute deviations from Gaussian expectation.

\begin{figure}
\begin{center}
\includegraphics[scale=0.62]{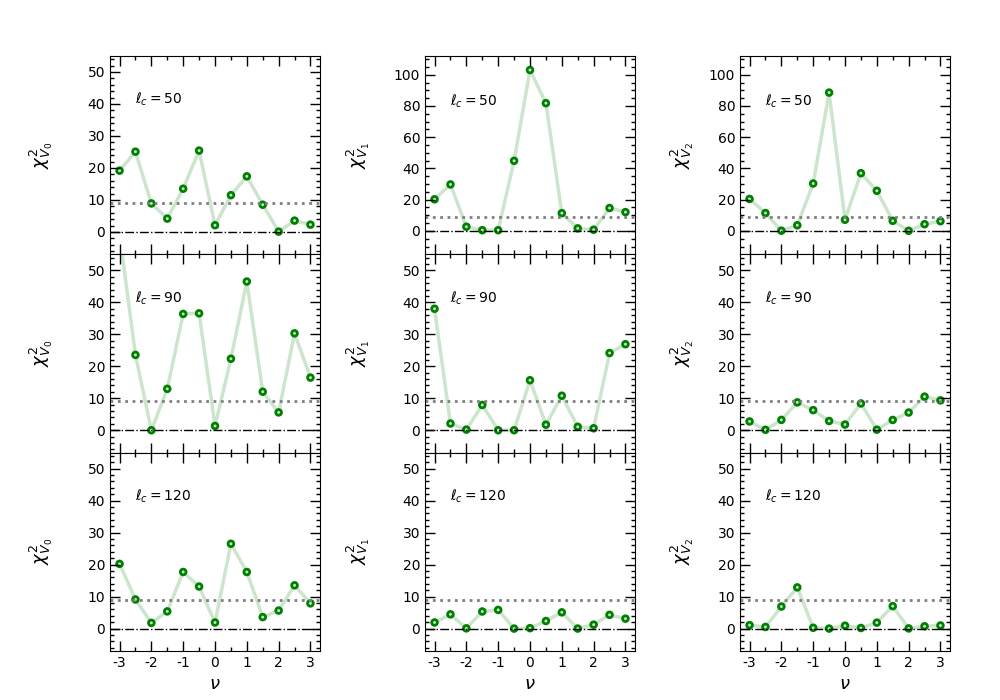}
\end{center}
\caption{\small $\chi^2_{V_k}$ is shown as a function of threshold for each scalar MFs for $u_c=25$ K, for different $\ell_c$. It is seen that $\chi^2_{V_k}$ gets close to 3-$\sigma$ for all thresholds as we increase $\ell_c$. This indicates that the level of non-Gaussianity decreases as we remove more bright  regions as well as large-scale structures in the Haslam map.} 
\label{fig:ChiSqr}
\end{figure}

%\textcolor{magenta}{
For each of the three scalar MFs, $\chi^2$ for the case of $u_c=25$ K are shown in figure \ref{fig:ChiSqr}. The line for $\chi^2=9$ which  corresponds to $3\sigma$ is shown by the black dotted line for reference.  Note that except for the case of $V_0$, the $y$-axis scales for the top panels showing $\ell_c=50$ are different from the lower panels. 

We observe that for all three MFs, $\chi^2$ values decrease towards $\ell_c=120$. The rate of decrease, however, varies. 
For $V_0$, the $\chi^2$ values are higher than 9 for most threshold values, at all  $\ell_c$. For both $V_1$ and $V_2$, the values of $\chi^2$ decrease as $\ell_c$ increases, and for $\ell_c=120$, are smaller than 3$\sigma$ for all threshold values (except at $\nu=-1.5$ where it is higher than 9 for $V_2$). 

We get the value of $\chi^2$ for $V_0$ averaged over all threshold values for $\ell_c=120$ is 11.08, corresponding to statistical significance of 3.3$\sigma$.  Therefore, from the behaviour of $V_0$, we conclude that the Haslam map for $u_c=25$ K is  mildly non-Gaussian with statistical significance  3.3$\sigma$ at the smallest scale probed in this work. The nature of the non-Gaussianity is of the kurtosis type given by the cumulant $K_0$ since the non-Gaussian deviation of $V_0$ is determined by $K_0$.

%%%%%%%%%%%%%%%%%%%%%%%%%%%%%%%%%%%%%%%%%%%%%%%%%%%%%%
\subsection{Statistical isotropy of the Haslam map}
\label{sec:iso}

Next, we discuss the results for $\mathcal{W}_1$. We again quantify the difference between the Haslam map and the Gaussian simulations as:
\begin{eqnarray}
  \Delta\mathcal{W}_1 &\equiv& \mathcal{W}_1^{\rm Haslam} - \mathcal{W}_1^{\rm (G)}, \\
  \Delta\alpha &\equiv& \alpha^{\rm Haslam} - \alpha^{\rm (G)},
\end{eqnarray}
where  $\mathcal{W}_1^{\rm (G)}$ and $\alpha^{\rm (G)}$ are the values obtained from the Gaussian simulations, while the superscipt `Haslam' refers to the values for the Haslam map. As in the previous case, the deviations $\Delta\mathcal{W}_1$ and $\Delta\alpha$ are normalized with $\mathcal{W}_1^{\rm G,max}$ and $\alpha^{\rm G,max}$, respectively. We compute them for each Gaussian isotropic simulation. 

\begin{figure}{t}
\begin{center}
\includegraphics[scale=0.655]{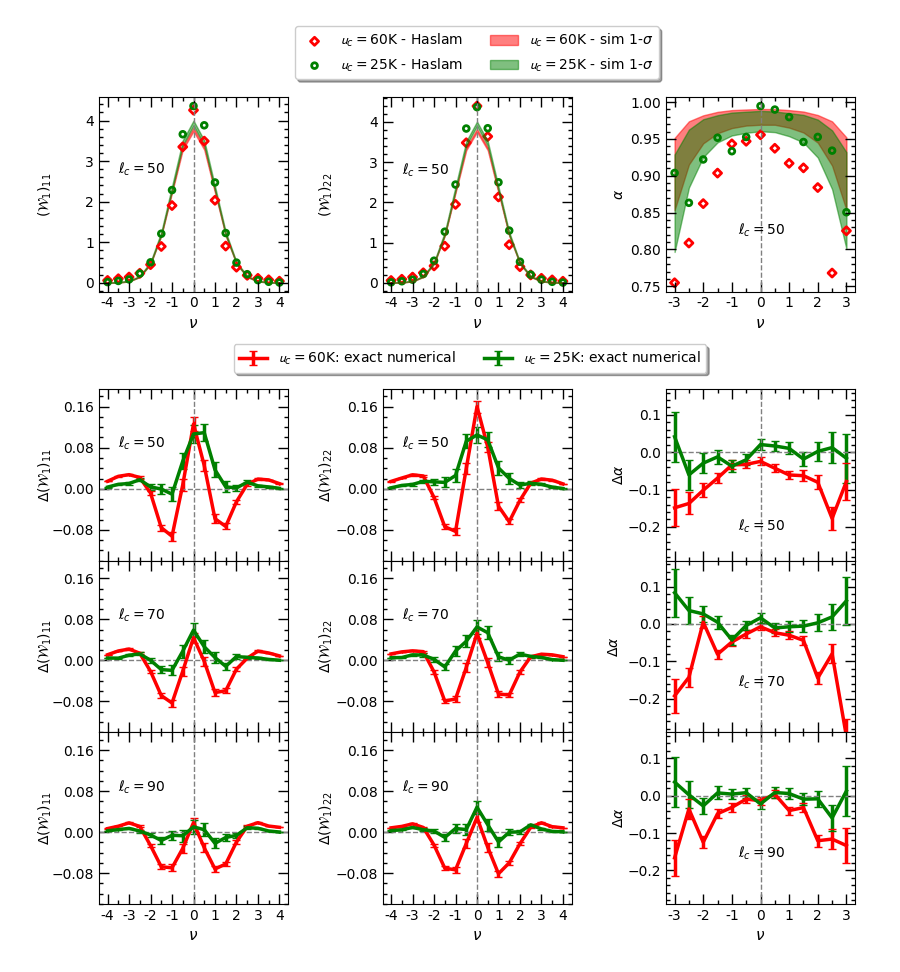}
\end{center}
\caption{\small The first row shows the two diagonal components of tensor MF, $\mathcal{W}_{1}$ and the anisotropy parameter, $\alpha$ for two temperature thresholds $u_c=60$ K and $u_c=25$ K (red diamonds and green circles, respectively) for $\ell_{c}=50$.  The ensemble mean and $1\sigma$ width from 1000 simulated maps are also plotted to show the deviation. Note that, for $\alpha$, the threshold range is (-3:3). The remaining rows represent the ensemble mean and $1\sigma$ width of the deviations, $\Delta \mathcal{W}_{1}$ and $\Delta \alpha$ with respective normalizations, for $u_c=60$\,K (red solid lines) and $u_c=25$\,K (green solid lines),  for $\ell_{c}=50,70,90$.
}
\label{fig:TMFs}
\end{figure}

The top row of figure~\ref{fig:TMFs} shows the diagonal elements of $\mathcal{W}_1$ (left and middle) and $\alpha$  for $u_c=60$  and $25$ K, for $\ell_c=50$ K. The colour coding is the same as in figure~\ref{fig:scalars}. The mean values obtained from the 1000 Gaussian isotropic simulations corresponding to each $u_c$ are shown by the solid lines, along with the 1$\sigma$ regions. 
The ensemble mean of the diagonal elements of $\Delta\mathcal{W}_1$, along with 1$\sigma$ error bars are shown in columns one and two of the lower panels, 
for the same values of  $u_{c}$ and for $\ell_{c}=50, 70, 90$. The physical information that can be deciphered from the elements of $\Delta\mathcal{W}_1$ is similar, up to statistical fluctuations, as $\Delta V_1$, and we find that the behaviour is the same, as expected.  
Ensemble mean and $1\sigma$ error bars of $\Delta\alpha$ are shown in the third column of figure~\ref{fig:TMFs} for the same values of $u_c$ and $\ell_c$.  We observe that $\Delta\alpha$ is smaller for smaller $u_c$ implying that the field becomes more isotropic when the warmer regions are excluded. This corroborates what we infer from visual inspection that high-intensity regions 
have large scale correlations that appear to be direction-dependent.

%%%%%%%%%%%%%%%%%%%%%%%%%%%%%%%%%%%%%%%%%%%%%%%%%%%%%%%%%%%%%%%%%%%%%%%%%%%

Next, to proceed with the quantification of the statistical significance of any deviation from SI of the Haslam map, we take into consideration the fact that the $\alpha$ statistic follows the Beta probability distribution given by 
\begin{equation}
P(\alpha)= \frac{\Gamma(a+b)}{\Gamma(a)\Gamma(b)}\alpha^{a-1} (1-\alpha)^{b-1}
\end{equation}
where $a > 0,\ b> 0$ are parameters that depend on the cosmological model~\cite{Prava:2021}. 
Due to this reason, 
for accurate quantification of the statistical significance of deviation from SI,  we use the median value of $\alpha$, denoted by  $\widetilde\alpha^{\,\rm (G)}$ obtained from the 1000 simulations along with the 95\% confidence interval.  We find that the mean and median values differ by less than 1\% at all threshold values. Let us denote the 95\% confidence interval  about the median by $[\delta_1, \delta_2]$. For each threshold, we  determine $\delta_1$ and $\delta_2$ such that they satisfy 
\begin{equation}
\int_{\widetilde\alpha^{\,\rm (G)}-\delta_1}^{\widetilde\alpha^{\,\rm (G)}} {\rm d\alpha}\,P(\alpha)= \int_{\widetilde\alpha^{\,\rm (G)}}^{\widetilde\alpha^{\,\rm (G)}+\delta_2} {\rm d\alpha}\,P(\alpha)=0.475. 
\end{equation}

 Let  $\Delta\widetilde\alpha \equiv \alpha^{\rm Haslam} - \widetilde\alpha^{\rm (G)}$.  Then, to quantify the statistical significance of $\Delta\widetilde\alpha $, we use
the variable $\widetilde\chi$ which is defined as follows:
{\renewcommand{\arraystretch}{1.5}
\begin{equation}
 \widetilde \chi = \left\{
  \begin{array}{c}
 \frac{\Delta\widetilde\alpha}{\delta_1}, \quad {\rm if} \ \Delta\widetilde\alpha<0, \\
        \frac{ \Delta\widetilde\alpha}{\delta_2}, \quad {\rm if} \ \Delta\widetilde\alpha>0.
\end{array} \right.
\end{equation}  
$\widetilde\chi$ reduces to the square root of the standard chi-squared statistic for the Gaussian case. $|\widetilde\chi|>1$ implies   $\alpha^{\rm Haslam}$ is outside the 95\% confidence interval and, hence, exhibits statistically significant deviation from the simulations. The sign of $\widetilde\chi$ contains useful information \textendash~if it is negative, it means more anisotropic, while positive values means more isotropic than the median value. 
$\alpha$ at neighbouring thresholds are uncorrelated if the threshold bin size is sufficiently large, and we have seen that our choice of $\Delta\nu=0.5$ is large enough.

\begin{figure}{H}
\begin{center}
\includegraphics[scale=0.6]{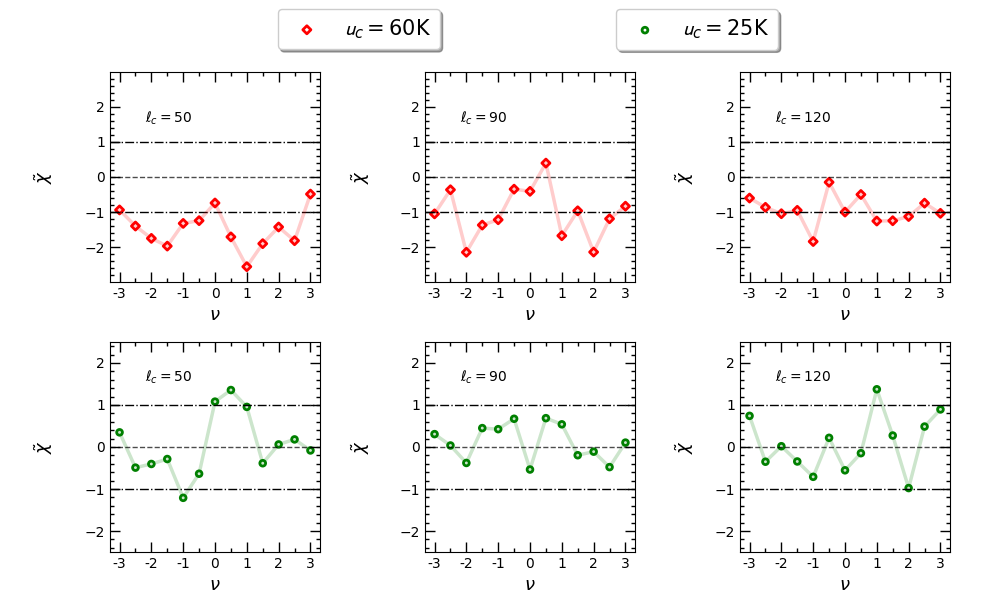}
\end{center}
\caption{\small $\widetilde\chi$ values of $\Delta\widetilde{\alpha}$ at each threshold values for $u_c=60$\,K (top panel) and $u_c=25$\,K (bottom panel), and for $\ell_c=50,90,120$. Lines corresponding to $|\widetilde\chi| = 1$ (95\% confidence interval) are marked for reference.
}
\label{fig:ChiSqr_alpha}
\end{figure}
Figure~\ref{fig:ChiSqr_alpha} shows $\widetilde\chi$ versus threshold values, for different $u_c$ and $\ell_c$. Lines corresponding to $|\widetilde\chi| = 1$ are marked for reference. 
For $u_c=60$ K, we find that $\widetilde\chi$ is negative for almost all threshold values for all $\ell_c$, which is due to  $\alpha^{\rm Haslam}$  being smaller than  $\widetilde\alpha^{\rm (G)}$. This implies that the Haslam map is genuinely anisotropic  in comparison to the isotropic simulations. Further, we observe that $\widetilde\chi$ becomes smaller as $\ell_c$ increases, indicating that the statistical significance of the anisotropy decreases at smaller angular scales. Hence, the smaller scale fluctuations of the field tend to follow isotropic distribution. 
For the case of $u_c=25$K, we see that $|\widetilde\chi| \le 1$ for most thresholds and the values fluctuate between positive and negative values for all $\ell_c$. 
Therefore, after excluding the warmer regions of the field (decreasing $u_c$), we find that the Haslam map exhibits isotropic behaviour even at relatively large scales.

%%%%%%%%%%%%%%%%%%%%%%%%%%%%%%%%%%%%%%%%%%%%%%%%%%%%%%%%%%%%%%%%%%%%%%%%%%%%
\section{Summary of results and their implications}
\label{sec:sec6}

Using scalar Minkowski functionals and Minkowski tensors, we have carried out careful investigation of the statistical properties of one of the major foreground components, namely the Galactic synchrotron given by the full sky 408 MHz Haslam map. 
The results are summarized as follows. 
\begin{itemize}
\item Firstly, we find that the overall level of non-Gaussian deviations does decrease as more high emission regions are masked, and as we go down to smaller scales. This is not a new result and corroborates findings in earlier works.

\item Our analysis reveals that the leading source of non-Gaussianity of the Haslam map, at all scales, arises from kurtosis terms, with skewness being sub-dominant. We demonstrate that in the cooler regions of the Haslam map, the non-Gaussian deviations of MFs agree very well with analytic perturbative expressions keeping up to kurtosis terms or second-order in the standard deviation of the field. 

\item The level of non-Gaussianity at the smallest angular scales of $\sim 1.5^{\circ}$ corresponding to $\ell_c=120$  probed by the Haslam map has a statistical significance of  3.3$\sigma$. This is determined by the area fraction, $V_0$, which has dependence only on one kurtosis cumulant, $K_0$. Hence, we  conclude that the assumption of Gaussian fluctuations in the synchrotron simulations is not appropriate at this scale. It is therefore important to analyse higher resolution synchrotron maps to determine the validity of the Gaussian approximation at scales smaller than the Haslam scale.    

\item Lastly, we test the statistical isotropy of the Haslam map and find that it becomes increasingly more isotropic in the cooler regions of the map as well as at smaller angular scales. This implies that the usual assumption of statistical isotropy at small scales in component separation methods is supported by the properties of the Haslam map.

\end{itemize}

It is interesting to note that the shape of non-Gaussian deviations of the MFs for the Haslam map is reminiscent of curvaton models of inflation where the leading contribution for non-Gaussianity comes from terms containing cubic self-coupling of perturbations with coupling parameter $g_{\rm NL}$ (see for example,~\cite{Enqvist:2008}). As a  consequence, we can expect that any residual Galactic synchrotron contamination in the CMB will predominantly bias  constraints on $g_{\rm NL}$.  Our results indicate that it may be possible to model the Galactic synchrotron fluctuations in the cooler regions, along the lines of inflationary perturbations, as an effective field that can be expanded as a Gaussian component plus a small perturbation of the type $\delta I(\vec x) \simeq \delta I^{\rm (G)}(\vec x)+g_{\rm NL}\left(\delta I^{\rm (G)}(\vec x)\right)^3$. Here, `G' stands for Gaussian component. 
We expect that $g_{\rm NL}$ can be related to  small scale fluctuations of the Galactic magnetic field and the distribution of relativistic cosmic ray electron and, hence, will be useful in constraining them. We will address this issue in the near future.

Our results also imply that any residual synchrotron component that contaminates the CMB will most likely not be captured by estimators of non-Gaussianity such as the bispectrum. Instead, it can be revealed by trispectrum or real space statistics such as MFs. Therefore, it is necessary to analyze foregrounds using a multitude of complementary statistics to uncover their true statistical nature. 
Further, it is important to probe non-Gaussianity and statistical isotropy of foreground fields at scales smaller than the resolution of the Haslam map. It is also important to probe different  Galactic components at frequencies ranging from radio to infra-red, such as done recently by  Coulton \& Spergel~\cite{Coulton:2019bnz} using the bispectrum.  We plan to carry out a detailed investigation of the non-Gaussian nature and SI of different foreground components at different frequencies relevant for the CMB, EoR and line intensity mappings using Minkowski tensors and trispectrum.

%%%%%%%%%%%%%%%%%%%%%%%%%%%%%%%%%%%%%%%%
%\section*{Acknowledgment}
\acknowledgments
    {We acknowledge the use of the \texttt{Nova} cluster at the Indian Institute of Astrophysics, Bangalore. 
    We have used \texttt{HEALPIX}~\cite{Gorski:2005,Healpix} and \texttt{Polspice}~\cite{Polspice,Chon:2004} packages to produce the results in this paper. The plots in this work are generated using \texttt{Matplotlib} library~\cite{Hunter:2007}. We thank A. Ducout and D. Pogosyan for giving us their code for the calculation of scalar Minkowski functionals.  We would like to thank T. Matsubara for useful communication. F.R. acknowledges support from the Department of Atomic Energy, Government of India, for visiting NISER, Bhubaneshwar, where a part of this work was carried out. F.R. would like to thank P. Goyal for helping out with the testing and analysis of Minkowski functionals. P.C. would like to thank K.~P.~Yogendran for useful discussion on the non-Gaussian nature of random fields. The work of P.C is supported by the Science and Engineering Research Board of the Department of Science and Technology, India, under the \texttt{MATRICS} scheme, bearing project reference no \texttt{MTR/2018/000896}. TG acknowledges support from the Science and Engineering Research Board of the Department of Science and Technology, Govt. of India, grant number \texttt{SERB/ECR/2018/000826}.  We thank the anonymous referee for the helpful comments and suggestions.}

%%%%%%%%%%%%%%%%%%%%%%%%%%%%%%%%%%%%%%%%%%%%%%%%%%%%%%%%%%%%%%%%%%%%%%%%%%%%

\appendix

\section{Consistency checks of the Gaussian simulations with Haslam data}
\label{sec:a1checks}

Gaussian isotropic simulations of Haslam 408 MHz all-sky map are obtained to quantify the statistics explored in this work. The power spectrum of the Haslam map is generated using  \texttt{PolSpice} \cite{Polspice,Chon:2004}. It gives the full angular power spectrum of any given map corrected for masking, beam and pixel effects and the residuals via the incomplete sky coverage. This power spectrum acts as input for generating Gaussian isotropic simulations using \texttt{HEALPIX} subroutines. In essence, the Haslam data and simulations are expected to match at the power spectrum level, and it is crucial to check the consistency of these simulations with respect to the data.

Using \texttt{anafast} subroutine of \texttt{HEALPIX} package, the pseudo power spectrum ($C_{\ell}$) is computed for Haslam data and 1000 simulations after following the pre-processing pipeline discussed in section \ref{sec:sec4}. In figure \ref{fig:Spectra}, the binned power spectra of the data as well as the mean $C_{\ell}$ and $1\sigma$ error bars from the simulations are shown for two of the masks ($u_c=60$\,K \& 25\,K)  used in our analysis . It is found that power spectrum for Haslam data and Gaussian isotropic simulations  match within  1$\sigma$ in both the cases. This confirms that the computation of Haslam power spectrum using  \texttt{PolSpice} is sufficiently accurate, and, therefore, the credibility of our results compared with respect to the generated Gaussian isotropic simulations.

\begin{figure}
\begin{center}
\includegraphics[scale=.6]{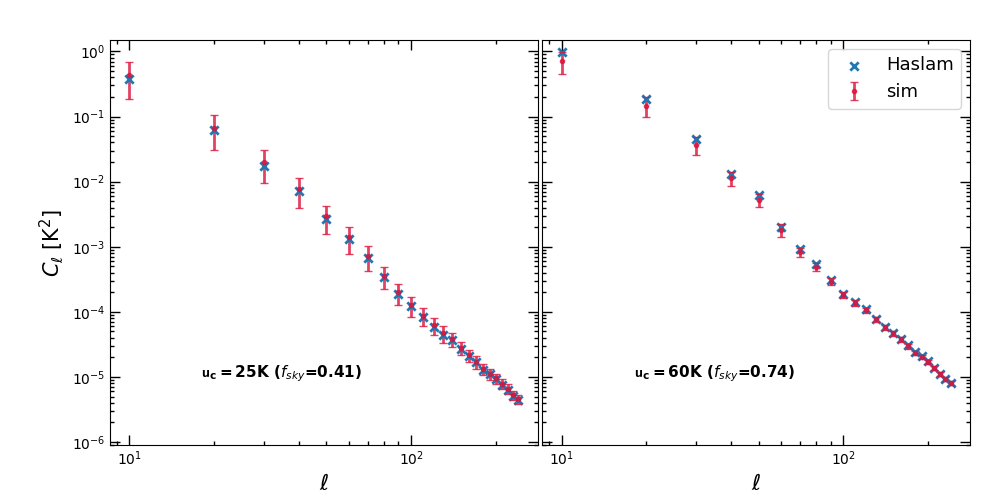}
\end{center}
\caption{\small The binned angular power spectum $C_{\ell}$ for the Haslam data and simulations obtained using \texttt{anafast} subroutine for two different sky fractions used in our analysis. The blue cross denotes the binned $C_{\ell}$ for Haslam data, while the red error bars indicates 1$\sigma$ width of binned $C_{\ell}$  from 1000 Gaussian isotropic simulations. It is seen that 1$\sigma$ matching is obtained for both the sky fractions. This confirms the accuracy of the power spectrum computed using \texttt{PolSpice} and, therefore, the credibility of our results compared with respect to the 1000 Gaussian isotropic simulations.}
\label{fig:Spectra}
\end{figure}

%%%%%%%%%%%%%%%%%%%%%%%%%%%%%%%%%%%%%%%%%%%%%%%%%%%%%%%%%%%%%%%%%%%%%%%
\section{Probability distribution function of the Haslam map}
\label{sec:pdf}

It is helpful to visualize the probability distribution function (PDF) of the Haslam map and the Gaussian simulations for different temperature and multipole cut values. To do so, we first define $\nu\equiv u/\sigma$ which is the  field value normalized by the $\sigma$ value, for each $\ell_c$. In figure~\ref{fig:pdf}, the  PDFs of the Haslam map, are shown as red diamonds for $u_c=60$ K and green circles for $u_c=25$ K, using the mask boundary threshold value, $s_m=0.9$. Visually, we find approximate symmetry about the field mean value zero, and decreasing levels of deviation with respect to the Gaussian expectations, with decreasing $u_c$ and increasing $\ell_c$. From this figure, we can anticipate that the kurtosis cumulants will have larger values compared to the skewness ones, supporting our observation that the nature of Haslam non-Gaussianity is predominantly sourced by kurtosis origin.

\begin{figure}%[!hbt]
\begin{center}
\includegraphics[scale=.55]{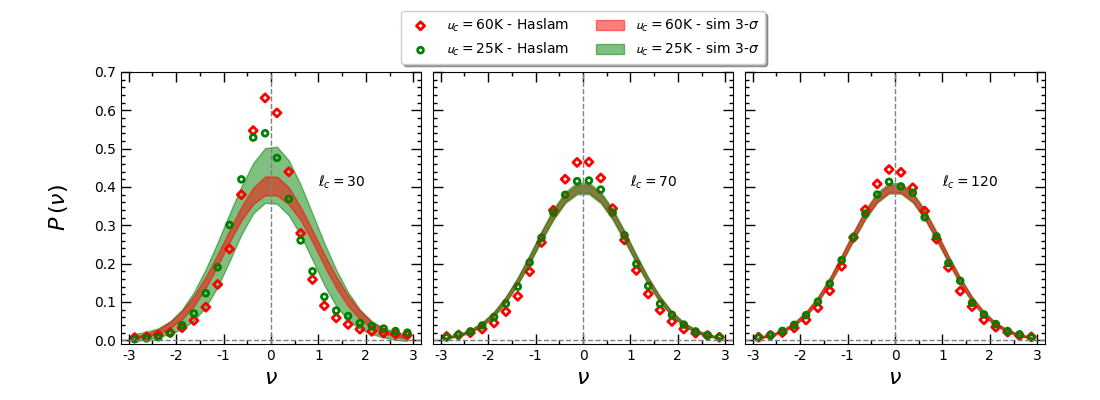}
\end{center}
\caption{\small The PDFs for the Haslam map and 1000 Gaussian simulations for different values of $u_c$ and $\ell_c$. The PDFs visually indicate approximate symmetry about the field mean value zero, and decreasing levels of deviation with respect to the Gaussian expectations, with decreasing $u_c$ and towards higher $\ell_c$. These results are obtained using the mask boundary threshold value, $s_{m}=0.9$.}  
\label{fig:pdf}
\end{figure}

\newpage

\section{Non-Gaussian deviations of Minkowski functionals for local type primordial non-Gaussianity }
\label{sec:a2fnlgnl}

\begin{figure}
\begin{center}
\includegraphics[scale=0.69]{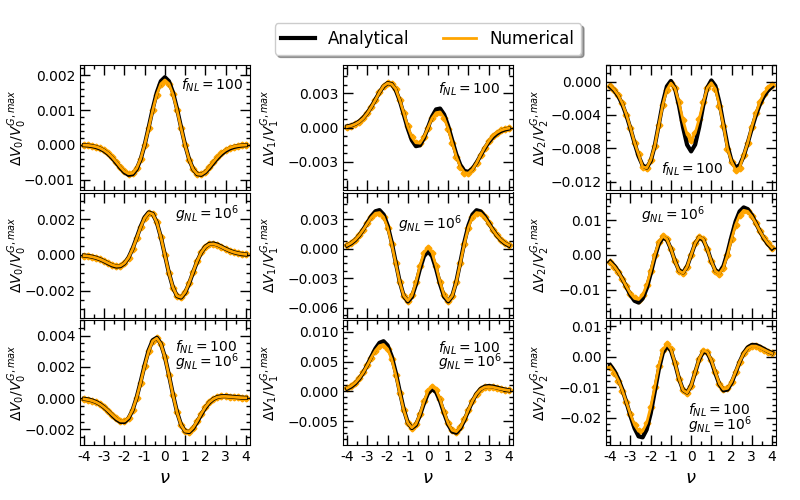}
\end{center}
\caption{\small The non-Gaussian  deviations for local type primordial non-Gaussianity for the cases $f_{\rm NL}=100,\, g_{\rm NL}=0$ (top), $f_{\rm NL}=0,\,g_{\rm NL}=10^6$ (middle) and $f_{\rm NL}=100,\,g_{\rm NL}=10^6$ (bottom) for the three Minkowski functionals. The black lines are the results obtained using the analytic formulae, and the orange lines are the results from the numerical calculations using method 1.} 
\label{fig:comparefnlgnl}
\end{figure}

Figure~\ref{fig:comparefnlgnl} shows the non-Gaussian deviations of the MFs for local type primordial non-Gaussianity parametrized by $f_{\rm NL}$ at first-order, and $g_{\rm NL}$ at second-order perturbations, as manifested in the CMB temperature. The results are in agreement with figure~5 of ~\cite{Hikage:2012}.  This figure serves to demonstrates the accuracy of the analytic formulae for non-Gaussian MFs in comparison with the numerical calculations for weakly non-Gaussian fields.  Moreover, it indicates that our analytic calculations for the skewness and kurtosis parameters, and numerical calculations for the MFs are correct.

%%%%%%%%%%%%%%%%%%%%%%%%%%%%%%%%%%%%%%

%\newpage
\bibliographystyle{JHEP}
\bibliography{haslamref}

 %%%%%%%%%%%%%%%%%%%%%%%%%%%%%%%%%%%%%%%%%%%%%%%%%%%%%%%%%%%%%%%%%%%%%%%%%%%%
\end{document}